\documentstyle[preprint,prd,aps]{revtex}

\begin{document}


\draft
\title{{\bf Remarks on the issue of time and complex numbers in canonical quantum gravity}}
\author{{\bf J.L. ROSALES }
\footnote{E-mail: rosales@phyq1.physik.uni-freiburg.de}\\
       {\em Fakult\"at f\"ur Physik,\\
         Universit\"at Freiburg,Hermann-Herder-Strasse 3 \\
        D-79104 Freiburg, Germany.}}
\date{\today}
\begin{center}
\maketitle
\begin{abstract}

Upon using some special  example in 
the homogeneous cosmological model
we develop the idea that, as a result of the arbitrariness of the factor
ordering in Wheeler-DeWitt equation, gauge phases can not, in general, 
being completely removed from 
the wave functional in  quantum gravity. The latter
may be conveniently described by
means of a remnant complex term in WDW equation depending on the factor 
ordering.  Taking this equation for granted we can obtain WKB complex
solutions and, therefore,
we  should be able to derive a semiclassical time parameter 
for the Schr\"odinger equation corresponding to matter fields in
a given classical curved space.

\end{abstract}
\end{center}

{\em Introduction.}

The precise formulation of quantum theory of gravity is not yet known.
However, such a theory should be developed in order to address some
fundamental questions in Physics like what happens at the Planck era of the
universe or during the final evolution of a black hole.

As approximate first step, we can cast general relativity into Hamiltonian
form and then, formally quantize it according to the canonical quantization
rules. The intention is to derive a wave equation analogously to the
Schr\"odinger picture of quantum mechanics; we thus obtain a set of
independent wave equations where the three-geometries and matter fields play
the role of configuration space. The main result of this approach is the
canonical hamiltonian constraint or Wheeler-DeWitt equation (WDW)
\cite{kn:DeWitt}

\begin{equation}
H\Psi(g^{3};\phi)=0 
\end{equation}

As a general rule, apart from a gauge fixing which would involve a global
arbitrary phase factor, the hamiltonian $H$ is supposed to be a real operator in
configuration space. 
An additional
observation is that time is absent and so the state of the universe (the
object to which we apply WDW equation) is stationary.

Yet, taking canonical quantum gravity for granted, we should 
understand the connection of this framework to more standard
concepts, particularly we must be able to derive the approximation
of Quantum Field Theory in curved Spacetime. The recovery of
this limit can be obtained by using an expansion, for the
{\em complex solutions} of (1)
in powers of the Planck mass (assuming that this represents a mass
far beyond the relevant energy scale for the matter fields). This is just a
Born-Oppenheimer approximation (see e.g.\cite{kn:Banks}). In recent years
\cite{kn:Kiefer4}, \cite{kn:Barbour}, \cite{kn:Smolin},
however, it has been notice that  there exist no sufficient
reason for the election of complex waves as the solutions of such a real
equation and, in general, we could speculate
about special prescriptions for the wave function\cite{kn:Nota}.
In this brief
report, on the other hand, we will try to show that the real feature of (1)
might be just a consequence of some fine tuning selection
corresponding to the operator factor ordering ambiguity of the theory 
and that, in some more general cases, arbitrary phases can not be completely
removed from the quantum theory. The latter would result in a remnant
complex term within the definition of the quantum Hamiltonian operator;
from this new tentative complex formulation of quantum gravity,
a  semiclassical time
would be implemented automatically after the direct
Born-Oppenheimer expansion. 

On the other hand, it has been claimed that
the mechanism of decoherence would account for an effective branching
of a given real solution into its complex components
\cite{kn:Kiefer2}\cite{kn:Halliwell},\cite{kn:Kiefer3},
which could perhaps be interpreted as a {\em symmetry breaking}
analogously to the case of chirality in sugar molecules \cite{kn:Zeh3}. This
would make possible, for an observer in a semiclassical universe, to detect
no interference whatsoever between the terms that appear in such a
superposition of waves. Finally, some coarse grained density matrix for the
relevant variables (e.g. the scale factor of the cosmological model) is
found to be {\em almost} diagonal after cosmological evolution (upon tracing
out gravitational fluctuations and matter fields which are not measured by
such a -classical- observer). This is possible since gravity always couples
to matter (which is now responsible for self-measurement of the universe).
Thus, in spite of the fact that Wheeler-DeWitt equation is invariant under
complex conjugation, the {\em actual states} should be intrinsically complex
and the whole process behaves as a phase transition in the cosmological
evolution. On the other hand, this argument by no
means avoids the possible presence of real superpositions in the quantum 
state (which
has not been measured since it represents the global information of the
state of the universe, including gravitational fluctuations and matter
fields). Only from this state we are allowed to derive the approximate
Schr\"odinger equation. Taking this
into account, we must, perhaps, consider some more direct possibilities in
order to justify the emergence of some semiclassical time parameter from the
asymptotic expansion about these very special complex WKB states.

{\em Quantum gravity in the homogeneous model.}

Our general discussion will be done for the minisuperspace approximation in
the homogeneous cosmological model. Therefore, as far as we are concerned on
the general features corresponding to this model, let us collect the general
properties of its well known classical and quantum theories \cite
{kn:Hawking1}, this model has the advantage of working on a one dimensional
space of variables for the set of Hamilton field equations and seems to be a
good approximation for the universe we observe.

In the homogeneous model of the universe we can also take into account the
coupling features of an scalar massive field $\phi $. It is defined by means
of the metric given by 
\begin{equation}
ds^2=-N(t)^2dt^2+a(t)^2d\Omega _3^2 
\end{equation}
$N$ is the lapse function, $a$ is the scale factor and $d\Omega _3^2$ is the
standard three-sphere metric. Expressing the scalar field as $(2\pi
^2)^{1/2}\phi $ with the quadratic potential $2\pi ^2m^2\phi ^2$ the action
is ($H_0^2=\frac \Lambda 3$,$\Lambda $ being the cosmological constant) 
\begin{equation}
S=-\frac 1{2\sigma ^2}\int dtNa^3\{\frac{\dot a^2}{N^2a^2}-\frac
1{a^2}+H_0^2-\sigma ^2[\frac{\dot \phi ^2}{N^2}-m^2\phi ^2]\} 
\end{equation}
Here, $\sigma $ denotes the Planck length in geometrodynamical units.
Varying the action with respect to $N$ lead to the constraint 
\begin{equation}
H=\frac{N}{2a^{3}}\{-a^{2}\sigma^{2}p_{a}^{2}+p_{\phi}^{2}+\frac{1}{%
\sigma^{2}} [H_{0}^{2}a^{6}-a^{4}]+m^{2}\phi^{2}a^{6}\}=0 
\end{equation}
where $p_{a} = -\frac{\dot{a}}{\sigma^{2}N}a$, $p_{\phi} = \frac{a^{3}}{N}%
\dot{\phi}$.

In the conformal gauge the hamiltonian is given by 
\begin{equation}
H=\frac{1}{2}\{-\sigma^{2}p_{a}^{2}+2u(a)\}+H_{m}(\phi,p_{\phi})
\end{equation}
where the matter Hamiltonian is given by 
\begin{eqnarray*}
H_{m}(\phi,p_{\phi})= \frac{1}{2}\{\frac{p_{\phi}^{2}}{a^{2}} + a^{4}m^{2}\phi^{2}\}
\end{eqnarray*}
and $u(a)=\frac{1}{2\sigma^{2}}(H_{0}^{2}a^{4}-a^{2})$.

It leads to
the Wheeler-DeWitt equation for the state functional of the universe written
as 
\begin{equation}
\{\sigma ^2a^{2-p}\frac \partial {\partial a}a^p\frac \partial {\partial a}-%
\frac{\partial ^2}{\partial \phi ^2}+V^2(a;\phi )\}\Psi(a;\phi)=0 
\end{equation}
where $V^2(a;\phi )=2a^2u(a)+a^6m^2\phi ^2$, is the {\em general relativity}
potential for the homogeneous model. The parameter $p$ takes into account
some of the factor ordering ambiguity of the theory. For the semiclassical
derivation of the Schr\"odinger equation for the matter fields the parameter 
$p$ is of little importance, however, following the suggestion of Dirac\cite
{kn:Dirac3}, we have to keep it in mind in order to define the correct
quantum theory, moreover, as we will see, the possibilities of the quantum
theory are not completely exhausted from the previous formulation.
In order to clarify this, let us proceed to discuss Dirac's
observation in the context of the homogeneous cosmological model.

We should start from assuming that the
quantum theory does not need to be a direct result of the classical
formulation we dispose and, therefore, we must take special care in
developing its physical possibilities.

In order to examplify our discussion we
could search for some special mini-superspace Hamiltonian
obtained from Legendre-transforming some Lagrangian coming
from the  addition in (3)  of some special gauge 
phase. The latter, of course, will always  preserve
classical equations of motion but, in general, would also
involve additional products of quantum (non-conmuting) operators
in such a new Hamiltonian. A relevant example of the previous idea
follows from selecting the Hamiltonian given by

\begin{equation}
H^c=-\frac{\sigma ^2}2p_a^2-\frac 1{2\sigma ^2}a^2-kp_aH_0a^2+H_m(\phi
,p_\phi ) 
\end{equation}
where $k$ is some real parameter and the product of non-conmuting
operators is shown explicitly. We have required the presence
of some cosmological constant, $H_{0}$, from dimensional considerations.
The linear term in the momentum would lead to the expected  complexification
of the constraint when expressing it quantum mechanically.

It is immediately obvious that, for $k=\pm 1$,
the difference between the Legendre
transformed Lagrangians corresponding to $H$ and $H^{c}$ ($%
L(H)=(p_{a}\partial/\partial p_{a}-1)H$) is just a total derivative 
\begin{equation}
L(H)=L(H^{c})+\frac{d}{dt}\alpha(a)
\end{equation}
where $\alpha(a)= -kH_{0}a^{3}/3\sigma^{2}$. This make spurious the choice
of the dynamics.
In this case (upon taking $k=1$), we get the constraint

\begin{equation}
\{\frac{\sigma^{2}}{2}a^{-p}\frac{\partial}{\partial a}a^{p}\frac{\partial}{
\partial a} -iH_{0}a^{2-q}\frac{\partial}{\partial a}a^{q}-\frac{a^{2}}{2
\sigma^{2}}+H_{m}(\phi,p_{\phi})\}\tilde{\Psi}=0 
\end{equation}

Since the physically meaningful quantity is $|\Psi|^{2}$
(or $|\tilde{\Psi}|^{2}$ ) we should develop the gauge transformation a
little bit further upon removing the resulting phase
\begin{equation}
\tilde{\Psi}=\Psi_{c}e^{-i\alpha(a)}
\end{equation}

The wave function $\Psi _c$ now satisfies, from (9) 
\begin{equation}
\{\sigma ^2a^{2-p}\frac \partial {\partial a}a^{p}\frac \partial {\partial
a}-\frac{\partial ^2}{\partial \phi ^2}+V_c^2(a;\phi )\} \Psi _c(a;\phi )=0 
\end{equation}
where 
\begin{equation}
V_c^2(a;\phi )=V^2(a;\phi )+i(p+2-q)H_0a^3
\end{equation}
and a complex term depending on the factor ordering is shown explicitly;
of course, it has not
classical interpretation and comes directly from the quantum theory.

We should take into account that (11) has only complex solutions. We are
allowed, therefore, to obtain directly WKB-like solutions and, thus, the
approximate Schr\"odinger equation for the matter field.
In order to achieve this let us write the wave function as follows
\begin{equation}
\Psi_{c}(a;\phi)=e^{iS(a)}\psi(a;\phi) 
\end{equation}
where $S(a)=\frac{S_{0}}{\sigma^{2}}+S_{1}+\sigma^{2}S_{2}+...$;
then, inserting (13) in (11) we obtain, for any $p$ and $q$, up to $%
O(\sigma^{0})$, the Hamilton-Jacobi equation for a matter free universe. 
\begin{equation}
(\frac{\partial S_{0}}{\partial a})^{2}+a^{2}-H_{0}^{2}a^{4}\approx 0 
\end{equation}
Therefore, as a result of the WKB regime of the solution the wave is
strongly correlated about particular classical configurations.

Also, $O(\sigma ^2)$ equations lead to 
\begin{equation}
i\frac{\partial \psi (a;\phi )}{\partial \tau }\approx H_m \psi
(a;\phi) 
\end{equation}
where we have defined an intrinsic time derivative 
\begin{equation}
\frac{\partial S_0}{\partial a}\frac{\partial }{\partial a}
\equiv \frac{\partial }{\partial \tau}
\end{equation}
Here, $\tau $ is nothing more than an affine parameter which labels the
points along the trajectories about which the wave function is peaked. In
order to define such an intrinsic semiclassical time we have not required
decoherence of macroscopic states in a real superposition since those wave
functions no longer satisfy equation (11).

Moreover, the only place where the complex factor is recovered is in $S_{1}$.
We obtain for it,
\begin{equation}
2\frac{\partial S_{1}}{\partial a}\frac{\partial S_{0}}{\partial a}= i\frac{
\partial^{2} S_{0}}{\partial a^{2}}+i \frac{p}{a}\frac{\partial S_{0}}{
\partial a}+i\gamma a
\end{equation}
where,$\gamma=(p+2-q)H_0$ is a new constant.

In order to understand the meaning of $\gamma$, we must, perhaps,
relate its very  origin from the conditions of normalizability of the
wave function in minisuperspace. To see this, let us assign a probability
meaning to $\Psi$ in the sense of Hawking and Page \cite{kn:Page}, i.e.,
$|\Psi|^{2}$ is proportional to the probability of finding a 3-surface $S$
with metric $h_{ij}$ and matter field configuration $\phi$; Moreover,
they suggested that $p=1$ solves a part of the factor ordering problem,
since, in the previous case, WDW equation becomes hyperbolic in configuration
space. On the other hand,  the behaviour of the semiclassical
wave functional at $a\ll H_{0}^{-1}$ is given, for any $\gamma$ by

\begin{equation}
|\Psi|^{2}\sim a^{-2}e^{-a^{2}/\sigma^{2}}
\end{equation}
therefore, the norm should be defined in terms of the following
finite quantity
\begin{equation}
<\Psi|\Psi>=\int da d\phi a^{2}\Psi^{*}(a;\phi)\Psi(a;\phi) <\infty
\end{equation}
and this is just possible for the semiclassical wave functions only in
case that $\gamma\neq 0$. This may be seen from the behaviour of
$|\Psi|^{2}$ at infinity derived from (17) and (14),i.e.,
\begin{equation}
|\Psi|^{2}\sim a^{-3 -\frac{\gamma}{H_{0}}}
\end{equation}
which does not depend on the value of the cosmological constant.
Now, from (19), if $\gamma$ would vanish exactly, the norm
$<\Psi|\Psi>$ diverges as the logarithm of the scale factor and, therefore,
we are not allowed to assign a probability amplitude meaning to $\Psi$.
The norm of the wave functional converges only  if $\gamma>0$. Thus,
we might undestand the very meaning of the new constant $\gamma$ from
the requirements of the quantum theory. On the other hand, $\gamma$ may
reach a value very close to zero but it should not be strictly zero.

Finally let us make some additional remarks indicating that there exits a
quantum spacetime governed by $\Psi_{c}$ and that the problem of obtaining
its classical properties still persists; this would lead to the mechanism of
self-measurement of the universe as established in \cite{kn:Kiefer5} and 
\cite{kn:Halliwell}, thus, only an effective classical universe is observed
since the different many branches described by the functional state $%
\Psi_{c} $ decohere. This is required since, even if one restrict attention
to a single WKB component of a general complex superposition, one can not
speak of a classical spacetime because WKB wave functionals are spread out
over all configuration space \footnote{%
I thank to Claus Kiefer for this comment}.

{\em Summary and conclusions}

One of the {\em Prima Facie} questions in quantum gravity concerns with the
meaning of its classical limit, i.e., how to construct the corresponding
quantum analogue of the classical theory of gravity we dispose\cite{kn:Isham}%
. In a practical sense the matter is usually decided on an {\em ad hominum}
basis. Here we have written down two specific proposals for the quantum
gravity theory of the universe i.e., it could be described either by a real
functional $\Psi$ as well as by, perhaps, its complex generalization given
by $\Psi_{c}$. We motivated the second possibility in order to recover some
semiclassical intrinsic time parameter in the Schr\"odinger equation for the
matter field. The state $\Psi_{c}$ might also contain new physics since the
complex potential has not classical interpretation. Nevertheless, we have
been more concerned on the appropriated specification of the semiclassical
limit in this quantum theory and, as far as this requires a superselection
rule which eliminates real superposition states, we have stressed the
selection of the complex state described by the solution of the complex wave
equation. Since this was found just from a very special minisuperpace, a
general analysis of the behaviour of this imaginary term in superspace is
still necessary.

{\em Acknowledgments}

The author wishes to express his thanks to David Hochberg, Jos\'e Luis
S\'anchez-G\'omez and Claus Kiefer for discussing the issue raised in this
paper. I also wish to express my warmest gratitude to Professor Hartmann
R\"omer in extending to me the hospitality of the University of Freiburg,
where part of this work was done. This work is supported by the Spanish
Ministry of Sciences scholarship number EX95 08960718.

\end{document}